\documentclass[reprint,amsmath,amssymb,aps,pdftex]{revtex4-1}

\usepackage{hyperref}
\usepackage{graphicx}    % to include figures
\usepackage{amsmath}    % great math stuff
\usepackage{amsfonts}    % for blackboard bold, etc
\usepackage{amsthm}    % better theorem environments.
\usepackage{bbm}
\usepackage[caption=false]{subfig}
\usepackage{dcolumn}% Align table columns on decimal point
\usepackage{bm}
\usepackage{amssymb}% http://ctan.org/pkg/amssymb
\usepackage{pifont}%

\newcommand{\bb}{\mathbf{B}}

\newcommand{\bimat}{\texttt{BiMAT }}
\newcommand{\matlab}{\texttt{MATLAB}\textsuperscript{\textregistered} }

\begin{document}

%\nocite{*}
\title{\bimat : a \matlab package to facilitate the analysis and
visualization of bipartite networks}

\author{Cesar O. Flores}
\email{Corresponding author: cesar7@gmail.com}
\affiliation{School of Physics, Georgia Institute of Technology, Atlanta, GA 30332}
\author{Timoth\'ee Poisot}
\affiliation{School of Biological Sciences, University of Canterbury, Private Bag 4800, Christchurch 8140, New Zealand}
\affiliation{D\'epartement de Biologie, Universit\'e du Qu\'ebec \`{a} Rimouski, 300 All\'ee des
Ursulines, Rimouski G5L 2C5, Qu\'ebec, Canada}
\affiliation{Qu\'ebec Centre for Biodiversity Sciences, Montr\'eal, Qu\'ebec, Canada}
\author{Sergi Valverde}
\affiliation{ICREA-Complex Systems Lab, Universitat Pompeu Fabra, E-08003 Barcelona, Spain}
\affiliation{CSIC-Institute of Evolutionary Biology, Universitat Pompeu Fabra, E-08003 Barcelona, Spain}
\author{Joshua S. Weitz}
\affiliation{School of Biology, Georgia Institute of Technology, Atlanta, GA 30332}
\affiliation{School of Physics, Georgia Institute of Technology, Atlanta, GA 30332}

\begin{abstract}

The statistical analysis of the structure of bipartite ecological
networks has increased in importance in recent years. Yet, both algorithms
and software packages for the analysis of network structure focus on
properties of unipartite networks. In response, we describe BiMAT,
an object-oriented MATLAB package for the study of the structure of
bipartite ecological networks. BiMAT can analyze the structure of networks,
including features such as modularity and nestedness, using a selection of
widely-adopted algorithms. BiMAT also includes a variety of null models
for evaluating the statistical significance of network properties. BiMAT
is capable of performing multi-scale analysis of structure - a potential
(and under-examined) feature of many biological networks. Finally, BiMAT
relies on the graphics capabilities of MATLAB to enable the visualization of
the statistical structure of bipartite networks in either matrix or graph
layout representations. BiMAT is available as an open-source package at
\href{http://ecotheory.biology.gatech.edu/cflores}{http://ecotheory.biology.gatech.edu/cflores}.

\end{abstract}

\maketitle

\section{Background}
Biological and social systems involve interactions
amongst many components.  Such systems are
increasingly represented as networks, where nodes
denote the interacting objects, and the edges
denote the interactions between them~\cite{newman2010networks}.
Of course, not all networks
are alike. For example, networks are often differentiated based
on whether or not individual nodes have the same types of
incoming and outgoing links.  A network is termed
unipartite if any node can potentially connect to
any other node, as in metabolic networks~\cite{jeong2000large},
food webs~\cite{cohen1978food,dunne2006network},
or friendships/contacts in a social network~\cite{watts1998collective}.
The interactions between nodes in such networks are often highly structured, i.e.
they differ from idealized networks in which the probability of interacting between
any two nodes is constant (i.e. the so-called Erd\"{o}s-Renyi graph~\cite{erdos1960evolution}).
Evaluating the structure of a unipartite network has spurred
the development of concepts such as modularity, small-world
structure, and hierarchy~\cite{newman2010networks}.
Measuring these structures has in turn, led to efficient implementations
of algorithms meant to quantify and characterize network structure,
primarily that of unipartite networks~\cite{bastian2009gephi,shannon2003cytoscape,
hagberg2008exploring,csardi2006igraph}.

\begin{figure*}
	\centering
	\includegraphics[width=0.75\textwidth]{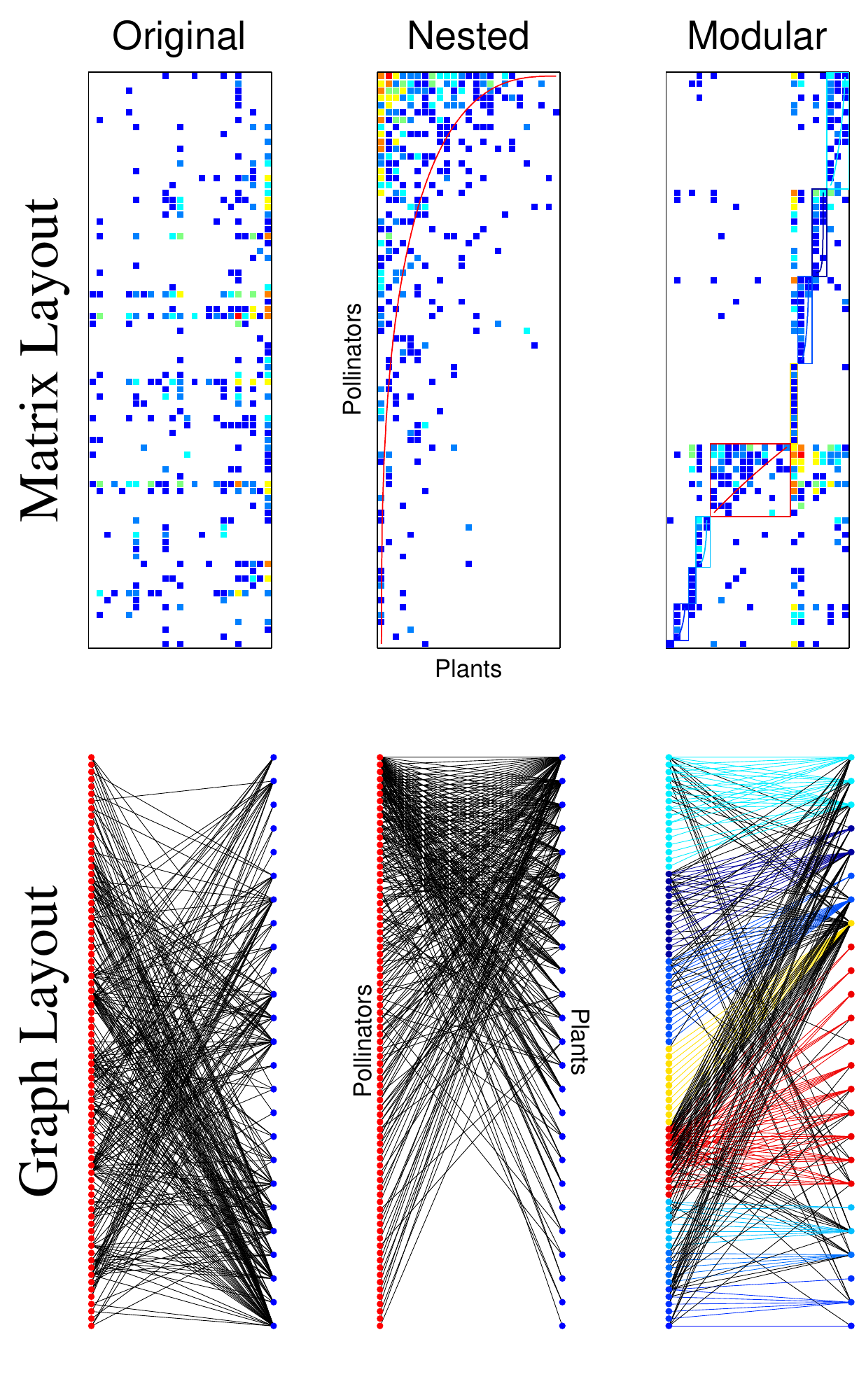}
	\caption{Schematic of an empirical bipartite network (plant-pollinator \cite{memmott1999structure})
	in matrix and graph layout using the original, nested and modular sorting
	of plant and pollinator nodes. 
	Color of cells are frequency of visits mapped to $\log$ scale, from small number of visits
	(darker blue) to large number of visits (dark red).
	While in the left panels no structure
	is apparent, the middle and right panels show the opposite. Through visual inspection of the panels, we may infer that the network is nested.}
	\label{fig.main_fig}
\end{figure*}

In contrast, a network is termed bipartite if nodes
represent two distinct types such that interactions can only 
occur between nodes of different types~\cite{Chartrand1985}.
The canonical example of bipartite networks is that of
interactions amongst
plant and pollinators, where links represent pollination~\cite{dunne2006}.
Indeed, an abundant literature has emerged
on the use of bipartite networks
and associated analysis techniques
for analysing plant--pollinators systems
\cite{Bascompte2007,Stouffer2011,Bascompte2003,Bastolla2009,Joppa2009}.
However, the concept of bipartite networks 
(and the specific methodology it carries) can be applied in different domains, including
the study of antagonistic networks such as
host-parasite 
interactions~\cite{PoisotBl2011, poisot2013structure,weitz2013phage,flores2011statistical, flores2012multi}.
Bipartite networks, like unipartite networks,
are rarely random in their structure, i.e. the probability of any potential link 
between each pair of nodes of 
different types is not equal.  Studies of both plant-pollinator
and host-parasite systems have shown that bipartite
networks can be (i) modular, i.e.
subsets of nodes often preferentially connect to each other,
rather than to other nodes~\cite{Olesen11122007}; (ii) nested, i.e. the interaction between nodes can be thought
of as subsets of each other~\cite{flores2011statistical, Bascompte2003}; (iii) multi-scale, i.e.
the structural properties of the network differ depending on
whether the whole or components are considered~\cite{flores2012multi}.
As an example, Figure \ref{fig.main_fig} shows Memmot \cite{memmott1999structure} plant-pollinator network, 
such that the nested and modular structure only becomes apparent when the appropriate sorting
is used.
Besides the importance of these metrics to quantity the structure of bipartite empirical data,
there is still not a self-contained library or software package
for analysing the structure of bipartite networks.

In response, we describe \bimat, an open-source software for
the analysis of bipartite networks. \bimat
is written in \matlab. Although \matlab is proprietary software, its
use has increased among ecological research groups due to the fact that producing results
and plots is easy and quick.
The library includes implementations of the most commonly
used algorithms for characterizing the extent to which  
a bipartite network exhibits modular, nested and multi-scale structure.
In addition to measuring the structure of a network,
\bimat also evaluates the statistical significance of this structure
given a suite of null models.  Finally, \bimat provides
a range of visualization tools for exploring bipartite network
structure in either matrix or graph layouts.
Here, we first describe the core definitions and methods
used in the analysis of bipartite networks.  Then, we describe the
implementation of \bimat and its application
to a number of examples drawn from virus-host interaction data.

\section{Methods}

\subsection{Bipartite ecological network}

A bipartite network, $\mathbf{B}$, is a network in which nodes can be divided in two sets $R$
(row nodes) and $C$ (column nodes) such that edges exist only across $R$ and
$C$. This type of network can be represented as a bipartite adjacency matrix
$\bb$ of size $m \times n$, where $m$ is the number of nodes in set $R$ and
$n$ is the number of nodes in set $C$.  In our implementation, $R$ and $C$ are
the node sets that are represented by the rows and columns of the bipartite
adjacency matrix in a \verb|Bipartite|
object.  Although \verb+BiMat+ takes quantitative matrices as input, all
algorithms implemented in \bimat first threshold these values such that
interactions are either present (1) or absent (0).
The number of links can be defined as $E = \sum_{ij} B_{ij}$.  Finally $k_i =
\sum_{j} B_{ij}$ and $d_j = \sum_{i} B_{ij}$ define the degree (number of interactions) of the two
kinds of nodes.

% The following table gives a summary of the basic properties
%in a network and the way they are called inside \verb+BiMat+ code.

%\begin{table}
%\caption{General properties of a bipartite ecological network}
%\begin{center}
%\begin{tabular}{l|l|r}
%	\hline
%	\hline 
%	Variable & Code & Description\\
%	\hline
%	$\bb$ & \verb+matrix+ & Bipartite adjacency matrix (boolean version)\\
%	$m$ & \verb+n_rows+ & Number of nodes in set $U$ \\
%	$n$ & \verb+n_cols+ & Number of nodes in set $V$ \\
%	$k_i = \sum_j B_{ij}$ & \verb|kk(i)| & Degree of node $i$ belonging to set $U$ \\
%	$d_j = \sum_i B_{ij}$ & \verb|dd(j)| & Degree of node $j$ belonging to set $V$ \\
%	$S=m\times n$ & \verb+size_matrix+ & Number of potential edges \\
%	$E=\sum_{ij} B_{ij}$ & \verb+n_edges+ & Number of edges \\	
%	$C=E/S$ & \verb+connectance+ & Connectance or fill \\
%	\hline
%	\hline
%\end{tabular}
%\end{center}
% \label{tab.isme.1}%
%\end{table}

\subsection{Algorithms}

\subsubsection{Modularity}

\bimat use the standard measure of modularity \cite{newman2006bmodularity}, which for
a bipartite network can be defined as (following Barber \cite{Barber2007}):
\begin{equation}
	Q_b = \frac{1}{E} \sum_{ij} \left( B_{ij} - \frac{k_i d_j}{E} \right) \delta(g_i,h_j),
	\label{eq.barber}
\end{equation}
where $g_i$ and $h_i$ are the module indexes of nodes $i$ (that belongs to set
$R$) and $j$ (that belongs to set $C$). The idea behind the last equation is
to maximize $Q$ by choosing the appropriate indexes for vectors $\mathbf{g}$
and $\mathbf{h}$.\ 
Significant debate concerns identifying
the optimal set of modules in the case of bipartite networks
\cite{fortunato2010community,sawardecker2009comparison}. 
In order to provide multiple options, 
\bimat uses three different algorithms to maximize Equation
\ref{eq.barber}: Adaptive BRIM \cite{Barber2007}, LP-BRIM
\cite{liuxin} and the leading eigenvector method \cite{newman2006bmodularity}.

\begin{itemize}
	\item \verb|AdaptiveBRIM|: 
The standard BRIM (for Bipartite
Recursively Induced Modules) algorithm works in the matricial notation version of
Equation \ref{eq.barber} given by:
\begin{equation}
	Q_b = \frac{1}{E}  \operatorname{Tr} \mathbf{R}^T \mathbf{\tilde{B}} \mathbf{T},
	\label{eq.qmat}
\end{equation}
where $\tilde{B}_{ij} = B_{ij} - \frac{k_i d_j}{E}$ is often called the modularity
matrix. Further, we replaced 
the delta function and vectors $\mathbf{g}$ and $\mathbf{h}$ by the $m \times c$ index matrix
${\bf R} = \left [{\bf r}_1 | {\bf r}_2 | ... | {\bf r}_m \right ]^T$ and the $n \times c$ index matrix
${\bf T} = \left [ {\bf t}_1 | {\bf t}_2 | ... | {\bf t}_n \right ]^T$, for row and column nodes, respectively, with $c$ denoting the number of modules \cite{Barber2007}.
Notice that nodes cannot be classified into more than one module. Hence, vectors ${\bf r}_i$ and
${\bf t}_i$ consist of a single one  (corresponding to the chosen module) with all the other 
entries being zero.  For example, 
$r_{ik} = 1$ if the $i$-th row node belongs to the $k$-th module with $r_{ij} = 0$ for all $j \ne k$.
Using the last expression, the standard BRIM algorithm
computes the optimal modularity by inducing the division of one
set of nodes (say vector ${\bf T}$) from the division in the other set of nodes (say vector ${\bf R}$).
At each step, BRIM assigns nodes of one type to modules in order to maximize the modularity.   
BRIM iterates this process until a local maximum is reached. 
However, the choice of a predefined 
number $c$ of modules limits the efficacy of the algorithm. Hence, we use an adaptive
heuristic \cite{Barber2007} to identify the optimal set of modules (and associated modularity $Q$).
This heuristic assumes that there is a smooth relationship between the number of modules
$c$ and the modularity $Q_b(c)$. For continuous and
 smooth landscapes, a simple bisection method ensures that we will find the optimal value of $c=c^\star$ 
 corresponding to maximum $Q_b$. Starting at $c=1$ (and modularity $Q_b(1)=0$ because all nodes belong to the same module)  the adaptive BRIM searches for optimal $c$ by repeatedly doubling the number 
 of modules while modularity  increases,  $Q_b(2c) > Q_b(c)$.  At some point, the search crosses a maximum in the 
 modularity landscape, i.e. $Q_b(2c) < Q_b(c)$,  and we interpolate the number of modules $c^\star$ to some
  intermediate value in the interval $(c/2, 2c)$. 
%
%
%
%\textbf{And BRIM is what exactly?}.  First, the number of modules $N$ is increased
%by a factor of two as long as the modularity value $Q_b$, as evaluated in
%a BRIM step given a fixed model number, continues to increase.  If
%modularity decreases, then a $N$ is selected via a
%bisection method between the two previously chosen module numbers.
\item \verb|LP&BRIM|: The algorithm is a combination between the BRIM and
LP (Label Propagation) algorithms. The heuristic of this algorithm consists in searching for the
best module configuration by first using the LP algorithm. This algorithm initially assigns
each node to a different module (label). At each interaction the module of each node is reassigned
to the module to which the majority of its neighbours belong to. The order
of node reassignment is random and ties are broken randomly.
The algorithm continues until convergence is achieved.
The standard BRIM algorithm
is used at the end to refine the results.

\item \verb|LeadingEigenvector|: This algorithm works with the unipartite adjacency matrix 
$\mathbf{A}$ of size $m+n \times m+n$ instead of the bipartite adjacency matrix $\mathbf{B}$.
The modularity using this notation can be defined for two modules in matrix notation 
as \cite{newman2006bmodularity}:
\begin{equation}
Q = \frac{1}{4E} \mathbf{s}^T \mathbf{\tilde{A}} s,
\end{equation}
where $\tilde{A}_{ij} = A_{ij} - \frac{k_i k_j} {2E}$ is the modularity matrix expressed
using the unipartite adjacency matrix with no distinction for degrees or rows and columns.
Further, for a particular division of the network into two modules, $s_i=1$ if node $i$
belongs to module 1 and $s_i=-1$ if it belongs to module 2. The idea of this algorithm is
that we can decompose the previous equation in a linear combination of
the normalized eigenvectors $u_i$ of $\mathbf{\tilde{A}}$ so that 
$\mathbf{s} = \sum a_i \mathbf{u}_i$ with $a_i=\mathbf{u}_i^T \cdot \mathbf{s}$:
\begin{equation}
Q=\frac{1}{4E}\sum a_i \mathbf{u}_i^T\mathbf{\tilde{A}} \sum a_j \mathbf{u}_j
   = \frac{1}{4E} \sum (\mathbf{u}_i^T \cdot \mathbf{s})^2 \alpha_i,
\end{equation}
where $\alpha_i$ is the eigenvalue of $\mathbf{\tilde{A}}$ corresponding to 
eigenvector $\mathbf{u}_i$.
The leading eigenvector name comes from the fact that in order to maximize the last equation
what we can do is to focus only in the sum term with the maximum eigenvalue $\mathbf{\alpha}_{max}$
which corresponds the leading eigenvector $\mathbf{u}_{max}$. This term can be maximized by trying
to maximize $\mathbf{u}_{max} \cdot \mathbf{s}$. Because $s_i$ can only have the values $\pm 1$,
this can be solved by assigning $s_i=1$ and $s_i=-1$ when $\mathbf{u}_{max_i} > 0$ and
$\mathbf{u}_{max_i} \le 0$, respectively, which completes the core of the leading eigenvector
algorithm. After performing the first iteration of the last process we will have a subdivision
of just two modules. Newman \cite{newman2006bmodularity} then explain that this process can
be applied recursively in each of the subdivisions. However, instead of isolating each subdivision
of each other, we apply this heuristic in the expression $\Delta Q$ which defines the change
of modularity that a new subdivision in an specific module will give us. The subdivision is only
accepted if $\Delta Q > 0$. For mode details about $\Delta Q$
we recommend to read \cite{newman2006bmodularity}. Finally, it is worth to mention that
in \bimat by default each subdivision is refined using the  Kernighan--Lin algorithm \cite{kernighan1970efficient} too.
The essence of this algorithm is swapping nodes between the two modules such that at each
step the node that gives the biggest increase in $Q$ or the smallest decrease (if increase is
not possible) is swapped. In a complete iteration all nodes are swapped with the constraint that
a node is swapped only once. The intermediate state during the iteration that has the biggest 
$Q$ is selected as the new configuration and the process repeats using this new configuration
until no improvement is possible. 
% 
%The modularity of a unipartite
%network can be expressed as a linear combination of the normalized 
%eigenvectors of the \textbf{modularity matrix -- what do you mean
%modularity matrix?}. In practice, the division of a network
%into modules is specified by first identifying the eigenvector with
%the largest positive real component of the 
%adjacency matrix. Next, all nodes whose projected values in the eigenvector
%are positive or negative are grouped into distinct modules.
%This procedure proceeds \textbf{how exactly... in an iterative fashion}.
%This algorithm only works for unipartite networks; hence
%\bimat first converts the bipartite matrix to its unipartite version.
\end{itemize}

%.Answering this
%question centers around about what is more important: the modularity value
%($Q$) or the module division ($\mathbf{g}$, $\mathbf{h}$). 
%In general, no
%unquestionable answer exist for this question. However, even if we accept that
%the best algorithm is the one that gives the highest $Q$, no algorithm can
%outperform the others in 100 \% of the cases.  Most of the time, however, the
%statistical results will not change by choosing a different algorithm. Finally,
%the authors suggest that depending on the size of the adjacency matrix the user
%can choose \verb|AdaptiveBrim| and \verb|LP&Brim| classes for small and large
%si.e. respectively; with the \verb|LeadingEigenvector| for the general case.
%
% This is your opinion... doesn't belong in the text

In addition to optimize the standard modularity $Q_b$ \bimat also evaluates
(after optimizing $Q_b$) an a posteriori measure of modularity $Q_r$
introduced in \cite{poisot2013posteriori} and defined as:
\begin{equation}
Q_r = 2\times\frac{W}{E}-1 %% This version returns values in 0;1
\end{equation}
where $W = \sum_{ij} B_{ij} \delta(g_i,h_j)$ is the number of edges that
are inside modules. Alternatively, $Q_r \equiv \frac{W-T}{W+T}$ where
$T$ is the number of edges that are between modules.  In other words, 
this quantity maps the relative difference of edges that are
within modules to those between modules on a scale 
from 1 (all edges are within modules) to $-1$ (all edges are between modules). This measure allows to compare the output of different algorithms.

\subsubsection{Nestedness}
Nestedness is a term used to describe the extent to which
interactions form ordered subsets of each other.
Multiple indices are available to quantify
nestedness (see \cite{ulrich2009consumer} for details about many of these
measures). Two of the most commonly used methods are:
NTC (Nestedness Temperature Calculator) \cite{Atmar1993,Rodriguez-Girones2006}
and NODF (for Nestedness metric based on Overlap and Decreasing Fill)
\cite{almeida2008consistent}. Both of these are implemented in \bimat
and are summarized below:

\begin{itemize}
	\item \verb|NestednessNTC| (NTC): A `temperature', $T$, of the interaction matrix
	is estimated by resorting rows and columns such that the largest quantity of interactions
	falls above the isocline (a curve that will divide the interaction from the non-interaction
	zone of a perfectly nested matrix of the same size and connectance). In doing so, the value of $T$ quantifies the extent to which interactions only take place in
	the upper left ($T\approx0$), or are equally distributed
	between the upper left and the lower right ($T\approx100$).
	Perfectly nested interaction matrices can be resorted
	to lie exclusively in the upper left portion and hence
	have a temperature of 0. The value of temperature
	depends on the size, connectance and structure of
	the network. Because the temperature value quantifies departures from perfect nestedness, we define
	the nestedness, $N_{NTC}$, of a matrix to range from 0 to 1,
	$N_{NTC}=(100-T)/100$, such that $N_{NTC}=1$ when $T=0$ (perfect nestedness) and
	$N_{NTC}=0$ when $T=100$ (checkerboard).
	\item \verb|NestednessNODF|: NODF is independent of row and column order. This algorithm measures the
	nestedness across rows by assigning a value $M_{ij}^{\text{rows}}$
	to each pair $i$, $j$ of rows in the interaction
	matrix\cite{almeida2008consistent}:
	\begin{equation}
		M_{ij}^{\text{rows}} = \begin{cases}
			0 & \text{if $k_i=k_j$}\\
         n_{ij}/\min(k_i,k_j) & \text{otherwise}
		\end{cases}
		\label{eq.Mrows}
	\end{equation}
	where $n_{ij}$ is the number of common interactions between them. A similar term
	is used for the column contributions, such that the total nestedness is defined as:
	\begin{equation}
		N_{NODF} = \frac{\sum_{i<j} M_{ij}^{\text{rows}} + \sum_{i<j} M_{ij}^{\text{columns}}}
		{m(m-1)/2 + n(n-1)/2}.
		\label{eq.nodf}
	\end{equation}
	However, \bimat redefined Equation \ref{eq.Mrows} (and its column version), such that the last equation can be more easily vectorized:
	\begin{equation}
		M_{ij}^{\text{rows}} = \frac{(\mathbf{r}_i \cdot \mathbf{r}_j) \delta(k_i,k_j)}
		{\min(k_i,k_j)},
	\end{equation}
	where $\mathbf{r_i}$ is a vector that represents the row $i$ of the bipartite adjacency
	matrix. Equation \ref{eq.nodf} can be rewritten in terms
	of adjacency matrix multiplications (see code for details). This new vectorized
	version of calculating the $N_{NODF}$ value outperforms the naive one (using loops)
	by a factor over 50 in most of the matrices that we tested.
\end{itemize}

Note that a new eigenvalue-eigenvector approach to evaluating nestedness
has recently been introduced
\cite{staniczenko2013ghost},
which will be introduced in a future \bimat release.

%Which algorithm to use is a decision that the user need to do based in his goals.
%Currently, the NODF is the one that is used in most of the recent nestedness publications.
%It has the advantage that it does not depend on the original sorting of the matrix
%and is a lot faster than the NTC algorithm. Further, the statistical test about
%nestedness significance becomes stronger when the NODF value is used. 
% Again, opinion, and very informal.  The point is that both are available, just leave as is.

%\subsection{Bi}
%
%
%For nestedness we use two different algorithms to differentiate in both complexity and
%kind of results. The first one is the NTC (Netedness Temperature Calculator) algorithm
%that depend on the distance of unexpected matrix cells to a predefined curve called
%isocline. The second one is the NODF algorithm which is based in the overlapping and
%decreasing filling across rows and columns in the matrix.

\subsection{Statistics}

\subsubsection{Null Models}

We propose four null models to test the
significance of measured nestedness and modularity (see
\cite{Bascompte2003,ulrich2007null,flores2011statistical,poisot2013structure}
for more details). These null models generate random networks 
through a Bernoulli
process, where the probability of interactions are determined following
different rules.  Define $k_i$ as the degree of a node $i$ of the column class and $d_j$ as the degree of a node $j$ of the row class.
Then, the probability
that two nodes (of distinct classes) interact, $P_{ij}$ is:
%\mathbf{Very hard to follow... "received by", "species", etc. Just re-write to make it simple and more general... If $\mathbf{V}$ and $\mathbf{G}$ are vectors with the number
%of interactions respectively received by lower level species, and established
%by upper level species, $L$ is the number of interactions in the network,
%and $l$ and $u$ are respectively the number of species at the lower and
%upper levels, then the probability of two species having an interaction in
%the random network, $P_{ij}$, is}

\begin{description}
 \item[EQUIPROBABLE], $P_{ij} = E/(mn)$ -- the connectance of the network is respected, but not the number of interactions in which each node is involved. 
 \item[AVERAGE], $P_{ij} = (k_{i}/n + d_{j}/m)/2$ -- the connectance, and the expected number of interactions in which each node is involved, are respected
 \item[COLUMNS], $P_{ij} = k_{i}/n$ -- the connectance, and the expected number of interactions of row nodes, are respected
 \item[ROWS], $P_{ij} = d_{j}/m$ -- the connectance, and the expected number of interactions of column nodes, are respected
\end{description}

By default, \bimat generate networks that can have disconnected nodes
(i.e. nodes with no edges to any other nodes in the network).
However the user can impose a constraint that all nodes must be
connected to at least one other node (if possible) in the null model
generating process.  Note that
\bimat does not include some of the most constrained null models, 
\textit{e.g.}, random networks that respect not only the expectation
of connectance and degree but also the \emph{exact} degree
sequences as the original network~\cite{staniczenko2013ghost},

\subsubsection{Statistic Values}

Once an ensemble of random networks is specified, \bimat 
will return the following values:

\begin{itemize}
	\item \verb|value|: value to be tested (\textit{e.g.} nestedness or modularity).
	\item \verb|random_values|: the values of all random replicates.
	\item \verb|replicates|: number of replicates used during testing.
	\item \verb|mean|: mean of the replicate values.
	\item \verb|std|: standard deviation of the replicate values (note that
distributions of network values are not necessarily well described by a normal distribution).
	\item \verb|zscore|: The $z$-score of \verb|value| assuming that the replicate values represent the
	entire population.
	\item \verb|percentile|: The percentage of replicate values that are smaller than \verb|value|.
\end{itemize}
% The reader should know what to do at this point.

\subsubsection{Extended statistics}
As described above, \bimat enables
the evaluation of the statistical significance of modularity and nestedness.
Additional statistical evaluation is possible, including
the capability to conduct a meta-analysis and a multi-scale analysis.

\begin{description}
	\item[Meta analysis]: \bimat can simultaneously analyse
the network structure of a  set of related
	bipartite networks (\textit{e.g.} plant-pollinator networks or virus-host interaction networks). In which case, the distribution of network properties
of the set of networks can be analysed (see example I in the
Examples section for more details).
	\item[Multi-scale analysis]: Individual modules need not always
be homogeneous.  Hence, \bimat offers functionality to evaluate whether
or not the network has different structures at different scales, \textit{e.g.},
the overall network may be modular, but individual modules may be nested (see example
II in the Examples section for more details).  
%Further,
%if nodes have additional metadata (e.g., location area, trophic level,
%phylogenetic classification), then \bimat can 
%examinne the correlation between modules and node classification.
\end{description}

%\section{Multi-Scale Analaysis}
%
%\bimat can perform a multi-scale analysis after finding the module configuration of
%the bipartite ecological network. By performing this analysis, the user can gain
%insight about the internal structure and multi-scale properties of a bipartite
%ecological network. Because the analysis needs of the module configuration of
%the network, it may be useful only when the network has a clear module division
%(high modularity), is big, and it is sparse.
%
%The multi-scale analysis consist of two steps:
%\begin{enumerate}
%	\item To perform the modularity and/or nestedness analysis inside the detected
%	modules. This will allow the user to see if internal modules present modularity
%	and/or nestedness even when the entire network as a whole does not present such
%	structure
%	\item To find if a correlation exist between the labels of rows and/or columns
%	and the module distribution of nodes.
%\end{enumerate}
%
%
%\subsection{Group Testing}
%
%\bimat is capable of performing statistical analysis across many matrices without need
%of interactions by the part of the user.

\section{The \bimat package}

\begin{figure*}
	\centering
	\includegraphics[width=\textwidth]{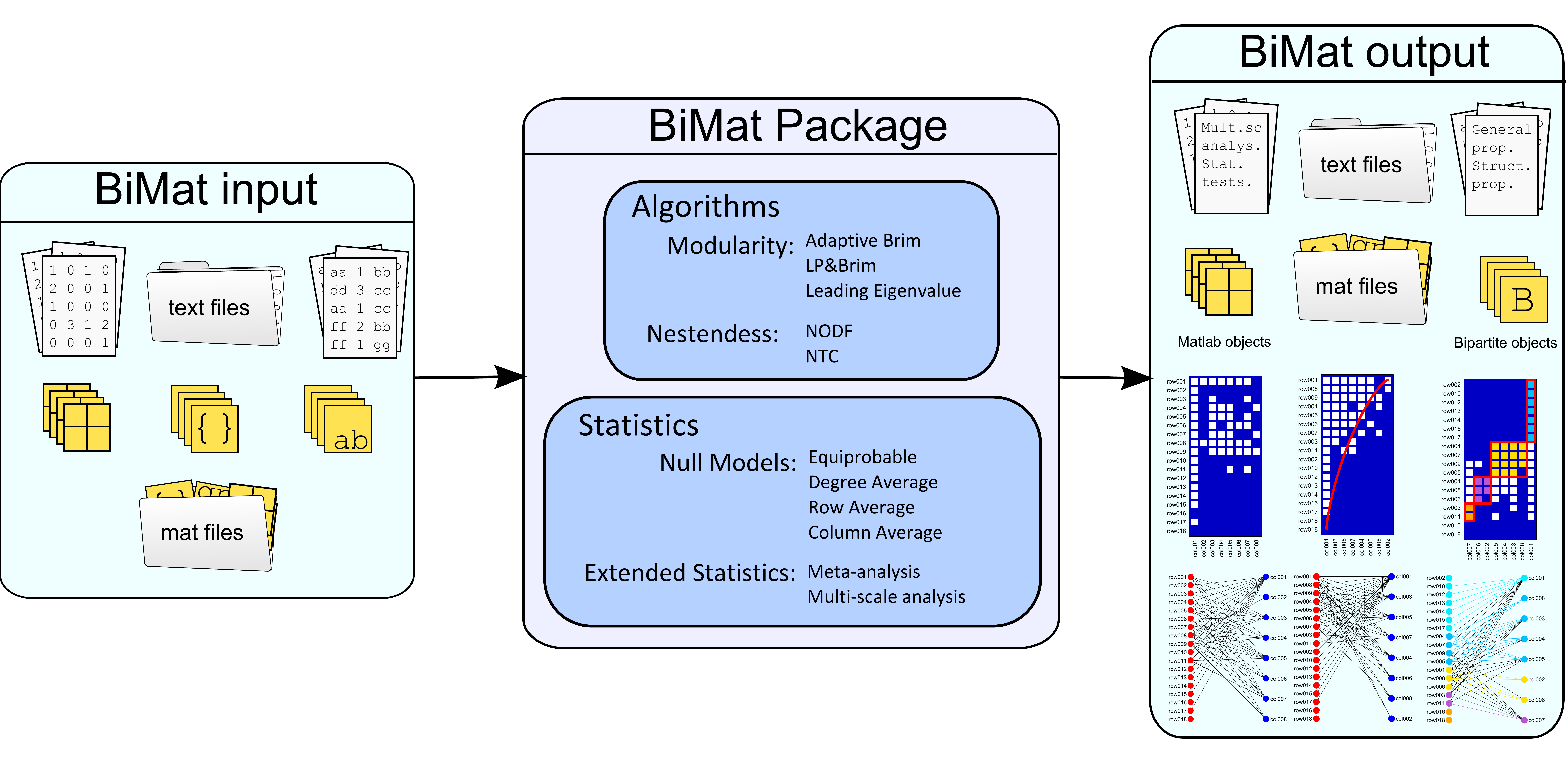}
	\caption{\bimat Workflow. The figure shows the main scheme of the \bimat package. \bimat can
	take matlab objects or text files as main input. The input is analysed mainly around
	modularity and nestedness using a variety of null models. The user may also perform an
	additional multi-scale analysis on the data, or if he have more than one matrix to perform
	a meta-analysis in the entire data. Finally, the user can observe the results via
	matlab objects, text files and plots.}
	\label{fig.workflow}
\end{figure*}

\bimat is a open-source package (see Figure \ref{fig.workflow}) written in \matlab. It is
primarily designed for the analysis and visualization of bipartite ecological
networks, thought it may be used for any type of bipartite networks. The
package aims to consolidate some of the
most popular algorithms and metrics for the analysis of bipartite ecological
networks in the same software environment. 
Specifically, the core features examined are bipartite
modularity~\cite{Barber2007} and nestedness~\cite{Atmar1993,almeida2008consistent}.
Further, \bimat include the necessary tools for analysing the statistical significance of
these values, together with tools for visualizing bipartite networks in such
a way that these properties becomes apparent to the user. 
\bimat utilizes an object-oriented framework which
enables users to extend the package.

\subsection{Usability}

Users are expected to be familiar with the \matlab environment.
However, \bimat has been designed so that even \matlab beginners or those with
very limited expertise can easily carry out a comprehensive analysis and
visualization of their data, in many cases with a single command.
Despite an emphasis on simplicity, \bimat still retains all of the
functionality and flexibility provided by the \matlab environment (\emph{e.g.},
all the results are returned to the current session workspace, the results
can be stored in \matlab files, and the class properties can be used for \matlab
plotting).  A complete start guide is distributed with the library.

%\begin{figure*}
%	\centering
%	\includegraphics[width=\textwidth]{figures/diagram_uml}
%	\caption{Class diagram of the \bimat package. This diagram shows the main classes of the \bimat
%	package together with the relationships among them. Bipartite class is the main one
%	inside the package and it is basically a bridge between all the functionality inside
%	\bimat.}
%	\label{fig.uml}
%\end{figure*}
%figure_main_mat

\subsection{Comparison with other software}
Current and popular available tools for the analysis of
complex networks include implementations that are predominantly: 
(i) visually oriented (\emph{e.g.} Gephi
\cite{bastian2009gephi}, Cytoscape \cite{shannon2003cytoscape}) or (ii)
library-package oriented (\emph{e.g.} networkx \cite{hagberg2008exploring}, iGraph
\cite{csardi2006igraph}). Unfortunately, these tools have a strong focus on the
analysis of unipartite networks, i.e. bipartite networks are treated as a
special case of a unipartite network. 
As a consequence, algorithms for the analysis of
unipartite networks, when applied to bipartite networks, are not intended to be optimal,
neither where designed to the study of ecological bipartite networks.
In contrast, 
specialized tools for the analysis of bipartite ecological networks
are available but they are very specific (\textit{e.g.}
ANINHADO \cite{guimaraes2006improving}, WINE \cite{galeano2009weighted}, and recently FALCON \cite{falconNest} focus only in nestedness analysis).

However, the authors acknowledge the existence of \verb|bipartite| \cite{bipartiteR}, a software library written in R. Thought this library initially included only nestedness analysis regarding internal
network structure, they just recently aggregated modularity analysis too \cite{dormann2014method}.
\bimat does not intent to replace this library but to complemented by bringing similar
tools to the \matlab ecology community. Further, \bimat also includes tools for the analysis of many
related networks (meta analysis) and for the analysis of different levels of the same network
(multi-scale analysis), which will facilitate the statistical analysis of bipartite ecological networks. Whereas \verb|bipartite| strives for exhaustivity, \bimat focuses on implementing a well-documented core of statistical procedures in an optimized way.

In summary, 
\bimat provides a broad selection of tools required for the analysis
and visualization of bipartite ecological networks.
As such, \bimat is aimed
towards empiricists seeking to apply a network perspective to their data,
and is particularly suited to exploratory analyses of data derived
from ecological, evolutionary, and environmental datasets. Table \ref{tab.software} show the current tools of current libraries.

\newcommand{\cmark}{\ding{51}}%
\newcommand{\xmark}{\ding{55}}%

\begin{table*}
\caption{Bipartite Ecological libraries}
\begin{center}
\begin{tabular}{r|c|c|c|c|c|c|c}
	\hline
	\hline 
	Software & Language & Open Source & Visualization & Nestedness & Modularity  & Meta-analysis & Multi-scale analysis\\
	\hline
	ANINHADO \cite{guimaraes2006improving}& Executable    & \xmark & \xmark & \cmark & \xmark & \cmark & \xmark \\ 
    WINE \cite{galeano2009weighted}       & \matlab/R/C++ & \cmark & \cmark & \cmark & \xmark & \xmark & \xmark \\ 
	FALCON \cite{falconNest}              & \matlab/R     & \cmark & \cmark & \cmark & \xmark & \xmark & \xmark \\ 
	\verb|bipartite| \cite{bipartiteR}    & R             & \cmark & \cmark & \cmark & \cmark & \xmark & \xmark \\ 
	\bimat                                & \matlab       & \cmark & \cmark & \cmark & \cmark & \cmark & \cmark \\ 
	\hline
	\hline
\end{tabular}
\end{center}
 \label{tab.software}%
\end{table*}

\subsection{Installation}

\bimat stable version can be downloaded directly from the main author webpage: 
\href{http://ecotheory.biology.gatech.edu/cflores}{http://ecotheory.biology.gatech.edu/cflores}. Last updated version can be downloaded from 
\href{https://github.com/cesar7f/BiMat}{https://github.com/cesar7f/BiMat}.

\subsection{License and bug tracking}

The software is distributed using FreeBSD license, which basically means that the user can redistribute
it, with or without modification for any kind of purpose as long as its copyright notices
and the licence's disclaimers of warranty are maintained. Though the license do not force users
to do so, we encourage them to cite this paper if the use of \bimat library leads to any
kind of scientific publication.

Users can report bugs directly in the github repository (see URL above), provided they
have a github account.

% I would rather have a github repo, with releases version. I can take care of that. It also allows to have Issue tracking, and a wiki.

% Also, it's important to specificy the code licence.

\subsection{Configuration}
The \bimat directory should be added to the \matlab paths.
At this point,
\bimat can be executed without any additional configuration.
The default parameters for algorithms
implemented in the \bimat package are available in
file \verb|main/Options.m|. 
Additional details are available in the Start Guide, including as part
of the \bimat package (and released here as Supplementary File X).

\subsection{Objected-Oriented Programming Scheme}
\bimat has been coded using the Objected-Oriented Programming (OOP) paradigm.
Note that understanding of OOP is not required for use of \bimat.
Nonetheless, the use of OOP is meant facilitate
maintainability and extensibility of the codebase. 
Access to \bimat functions is granted (with the exception of some static
classes) using instances of the class that implements the functions. 
%The class diagram with the main functions is depicted in \ref{fig.uml}.

The main package class is the \verb|Bipartite| class, whose only function is to
work as a common interface to all of the available statistical, algorithmic,
plotting, and input/output classes. Because of this OOP design pattern,
most of the \matlab functionality will be granted using the following syntax:
\begin{verbatim}
bip.class_instance_in_bip.method_name(arguments)
\end{verbatim}
\noindent
where \verb|bip| is a \verb|bipartite| instance created by the user,
\verb|class_instance_in_bip| is a property of the \verb|bipartite| class
which represents an instance of the class which has access to the method
\verb|method_name|. The method that is called will frequently have
direct read and writeable access to other properties inside \verb|bip|. Table
\ref{tab.oop.functions} shows the main calls from the Bipartite object,
assuming that the user call its bipartite instance \verb|bip|.

Note that the OOP capabilities of \matlab are not as extensive as those of OOP focus
languages (\textit{e.g.} python, Java, C++). As such, certain behaviours
have been emulated in \bimat, \emph{e.g.} static classes were emulated using
private constructors.
However, in contrast to other languages that enable OOP, 
\matlab enables users to
store created instances as \matlab objects in files.
This ensures that users can save, and subsequently load, 
the results of partial 
analysis.
%\footnote{We caution that the \bimat directory
%must be in the \matlab path in order to ensure that objects
%are loaded and that subsequent save commands include objects
%containing partial or complete analysis.}.
% I don't know anything about matlab, but it seems either obvious to users, or too technical for the paper

\begin{table*}
\caption{Some useful calls using the OOP approach}
\begin{center}
\begin{tabular}{l|l|l}
	\hline
	\hline 
	Call & Class & Description\\
	\hline
	bip.community.Detect() & BipartiteModularity & Calculate Modularity\\
	bip.nestedness.Detect() & Nestedness & Calculate nestedness\\
	bip.statistics.DoCompleteAnalysis() & StatisticalTest & Executes the required commands in order to have\\ 
	& & a complete analysis of nestedness and modularity\\
	bip.statistics.DoNulls() & StatisticalTest & Create the null model matrices\\
	bip.statistics.TestCommunityStructure() & StatisticalTest & Perform the statistical test for modularity values\\	
	bip.statistics.TestNestedness() & StatisticalTest & Perform the statistical test of nestedness value\\
	bip.internal\_statistics.TestDiversityRows() & InternalStatistics & Perform diversity analysis across rows\\
	bip.internal\_statistics.TestDiversityColumns() & InternalStatistics & Perform diversity analysis across columns\\
	bip.internal\_statistics.TestInternalModules() & InternalStatistics & Perform an statistical test\\
	& & for modularity and nestedness inside modules\\
	bip.plotter.PlotMatrix() & PlotWebs & Plot a matrix layout of the original data\\
	bip.plotter.PlotModularMatrix() & PlotWebs & Plot a matrix layout of the modular sorted data\\
	bip.plotter.PlotNestedMatrix() & PlotWebs & Plot a matrix layout of the nested sorted data\\
	bip.plotter.PlotGraph() & PlotWebs & Plot a graph layout of the original data\\
	bip.plotter.PlotModularGraph() & PlotWebs & Plot a graph layout of the modular sorted data\\
	bip.plotter.PlotNestedGraph() & PlotWebs & Plot a graph layout of the nested sorted data\\					
	\hline
	\hline
\end{tabular}
\end{center}
 \label{tab.oop.functions}%
\end{table*}

\subsection{Input/Output}
The class \verb|bipartite| is the main class of the package.
Hence, a user will usually need to work with at least
one instance of this class.  An instance of this class
requires a boolean \matlab matrix object,
representing the bipartite adjacency network.
Alternatively, a \verb|integer| matrix can be provided e.g.,
when the values represent categorical levels of interactions, and
these categorical levels can be included in the visualization tools.
Optional arguments
that can be passed are the row and column node labels and classification
classes.  These arguments need to be passed directly to the properties of
the \verb|Bipartite| object.  In practice, an object
of the class \verb|Bipartite| can be created as follows:
\begin{verbatim}
bip = Bipartite(matrix);
bip.row_labels = rowLabels; 
bip.col_labels = colLabels;
bip.row_class = rowClasses; 
bip.col_class = colClasses;
\end{verbatim}
in which the variables \verb|matrix, rowLabels, colLabels, rowClasses|
and \verb|colClasses| are previously defined variables.
Network information, including adjacency matrix and node labels,
can be read directory from data files using the
static class \verb|Reading|:

\begin{itemize}
	\item \verb|bip = Reader.READ_BIPARTITE_MATRIX(filename)|: The file should
        be in the following format:
	\begin{verbatim}
	1 0 0 2 0 0 0
	1 2 0 0 0 2 1
	1 1 0 0 1 2 1
	1 2 3 0 0 1 1
	2 1 1 1 0 0 0
	\end{verbatim}
Each row in the file represents 
a different outgoing set of interaction from a node (in set A) to
        a different set of nodes (in set B) in the columns. 
	All values different from 0
	are counted as interactions, such that evaluation of network structure
        utilizes Boolean information whereas visualization can leverage the non-negative
        strengths of interactions:
	\item \verb|bip = Reader.READ_ADJACENCY_LIST(filename)|: The file should be
        an ordered list of triples:
	\begin{verbatim}
	row_label_1 1 col_label_1
	row_label_1 1 col_label_2
	row_label_1 2 col_label_3
	row_label_3 1 col_label_2			
	row_label_3 3 col_label_1
	row_label_2 3 col_label_2	
	\end{verbatim}		
	such that the first and third columns represent
	nodes from sets A and B, respectively, and (an optional) second column 
        denoting the strength of interactions.
\end{itemize}

\subsection{Functional alternative}

Static functions can be used
as an alternative to interacting with the \bimat package in an OOP framework.
For example, the network can be visualized in a graph
or matrix layout as follows:

\begin{verbatim}
PlotWebs.PLOT_MATRIX(matrix);
PlotWebs.PLOT_GRAPH(matrix);
\end{verbatim}
instead of:
\begin{verbatim}
bp = Bipartite(matrix);
bp.plotter.PlotMatrix();
bp.plotter.PlotGraph();
\end{verbatim}

Table \ref{tab.func.functions} shows some of the most important static
functions that provide access to part of the
\bimat functionality.

\begin{table*}
\caption{Useful calls in the functional approach}
\begin{center}
\begin{tabular}{l|l}
	\hline
	\hline 
	Call & Description\\
	\hline
	BipartiteModularity.ADAPTIVE\_BRIM(matrix) & Calculate the modularity values using the Adaptive BRIM algorithm\\
	BipartiteModularity.LP\_BRIM(matrix) & Calculate the modularity values using the LP\&BRIM algorithm\\
	BipartiteModularity.LEADING\_EIGENVECTOR(matrix) & Calculate the modularity values using the Leading Eigenvector algorithm\\
	Nestedness.NODF(matrix) & Calculate the NODF values\\
	Nestedness.NTC(matrix) & Calculate the NTC values\\
	PlotWebs.PLOT\_MATRIX(matrix) & Plot the data in matrix layout\\
	PlotWebs.PLOT\_NESTED\_MATRIX(matrix) & Plot the nested sorted data in matrix layout\\	
	PlotWebs.PLOT\_MODULAR\_MATRIX(matrix) & Plot the modular sorted data in matrix layout\\			
	PlotWebs.PLOT\_GRAPH(matrix) & Plot the data in graph layout\\	
	PlotWebs.PLOT\_NESTED\_GRAPH(matrix) & Plot the graph sorted data in matrix layout\\	
	PlotWebs.PLOT\_MODULAR\_GRAPH(matrix) & Plot the graph sorted data in matrix layout\\			
	Printer.PRINT\_GENERAL\_PROPERTIES(matrix) & Print to screen the general properties of the network\\
	Printer.PRINT\_STRUCTURE\_VALUES(matrix) & Print the modularity and nestedness values of the network\\
	\hline
	\hline
\end{tabular}
\end{center}
 \label{tab.func.functions}%
\end{table*}

\subsection{Plotting}

The class \verb|PlotWebs| provides the required functions to visualize a bipartite network
in a matrix or graph layout. Visualization can utilize
(i) the original sorted version of the data, (ii) the nested sorted version of the data, or
and (iii) the modular sorted version of the data.
% Further these plots are not only
%visualized using the \matlab framework, but also can be directly stored in 
%\verb|eps|, \verb|jpg|, and \verb|pdf| files by the use of static functions inside the 
%plot class. 
\bimat represents the interaction data with colored cells when a matrix
layout is used. Rows and columns denote members of the
the two sets and cells denote interaction strength.
The format
of the matrix is specified by modifying the \verb|PlotWebs| class properties
before calling the plotting functions.  Further, the format of the matrix
will depend on what kind of sorting is used.  For example, the
modular sorting plot can color the cells according to the module to which
they belong to or the type of interaction.  Some features 
are restricted to particular sortings, \textit{e.g.}, plotting an isocline (see Methods)
is available only in the nested and modular sorting. 
Alternatively, the \verb|PlotWebs| can plot the data in a graph layout in which 
members of the two sets A and B are depicted using a stacked
set of circles to the left and right, respectively.
Lines are draw between sets that
interact. As for matrices, many of the properties of 
\verb|PlotWebs| can be used to specify the format of the plot (see documentation).
%. For example, 
%the size of the circles, the vertical and horizontal distance between circles (sets),
%the width of the interacting li.e. and so on.

In addition to this main class, \bimat has an additional plot class called \verb|MetaStatisticsPlotter| that
is used for plotting meta-analysis results (analysis of many networks). This class can plot
statistical results of the structural quantities of the algorithms, together with visual graph
or matrix layout representations of the networks (see Examples section).

%\subsection{Extensability}
%Given that \bimat has an OOP design, extend 
% (\emph{e.g.}, adding another
%modularity algorithm is as easy as write a \verb+DetectMethod()+ function).
%PLEASE EXPAND... give an example... perhaps?
%f
\subsection{Performance}

The \bimat packages leverages optimization tools of \matlab. 
For example, algorithms implemented in \bimat were vectorized to
improve performance.
In addition, a version of \bimat that uses the \matlab Parallel Computing Toolbox can be requested
to the corresponding author.

\section{Examples}
We present here two examples to illustrate the potential use of \bimat
for visualization and analysis of bipartite complex networks: (i) a meta-analysis of 38 different phage-bacteria interaction networks;
(ii) a multi-scale analysis of the largest phage-bacteria interaction network.
Scripts and data for these examples are included in the \bimat release
and additional documentation is included in the start guide.

\subsection{Example I: Meta-analysis}
The study of virus-host interactions includes examination of whom infects
whom.  Exhaustive assays of cross-infection of a set of phages
(viruses that infect and kill bacteria) and a set of bacteria 
are generally reported as a bipartite cross-infection matrix. 
These matrices can be standardized such 
that rows and columns represent bacteria and phages, respectively. 
The cell enrties in these matrices represent the level of infection between
phages and bacteria.   In a previous study,
Flores et al \cite{flores2011statistical} re-examined
38 such networks extracted from the published literature between 
1950 and 2011.  In doing so, the authors found that
phage-bacteria infection networks (as published) tend to be nested
and not modular.  \bimat can reproduce 
these results using the \verb|MetaStatistics| module.  

\begin{figure*}[!ht]
 \subfloat[Modularity statistics\label{subfig-1:dummy}]{%
  \includegraphics[width=0.47\textwidth]{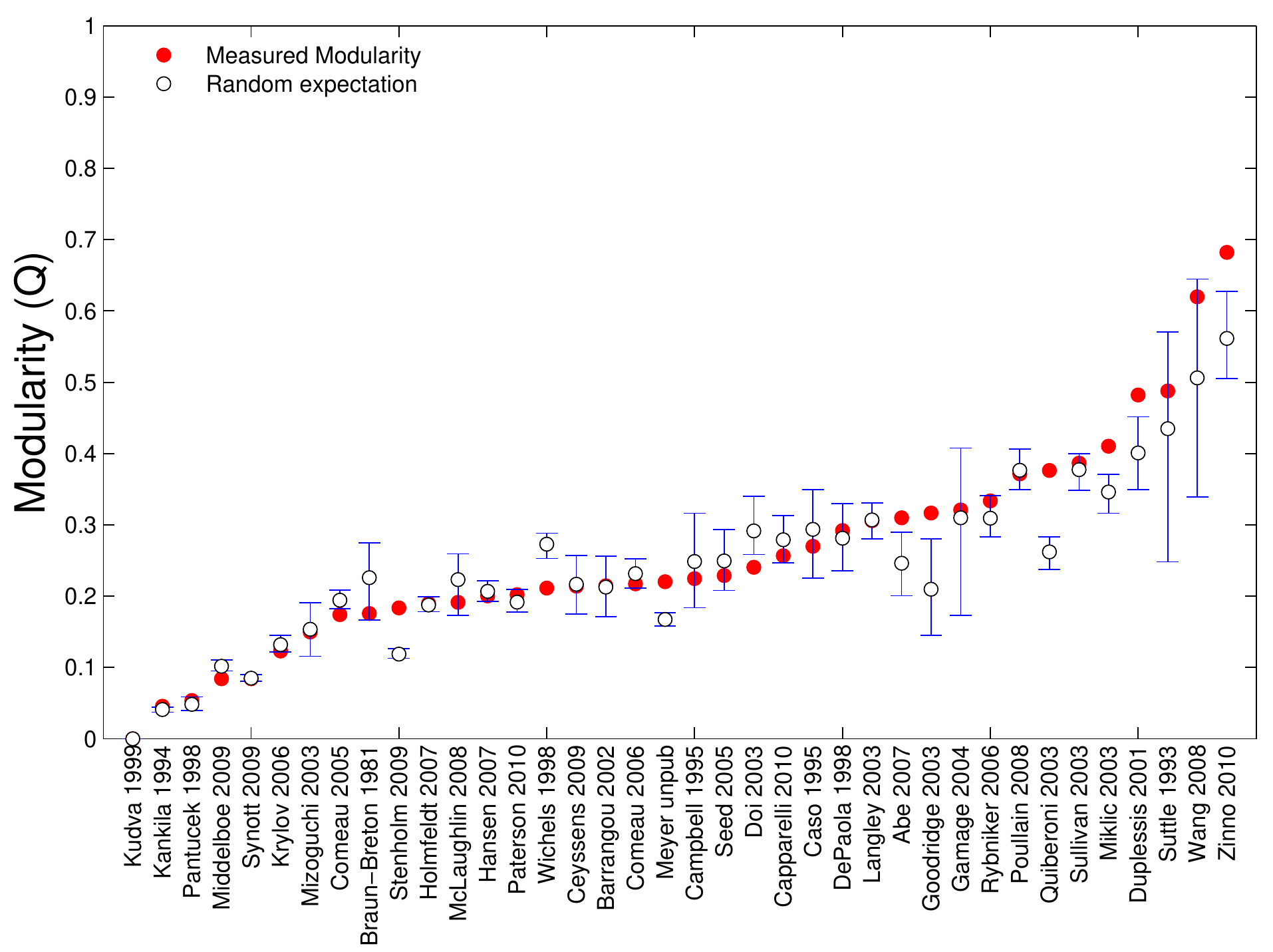}
 }
 \hfill
 \subfloat[Nestedness (NTC) statistics\label{subfig-2:dummy}]{%
  \includegraphics[width=0.47\textwidth]{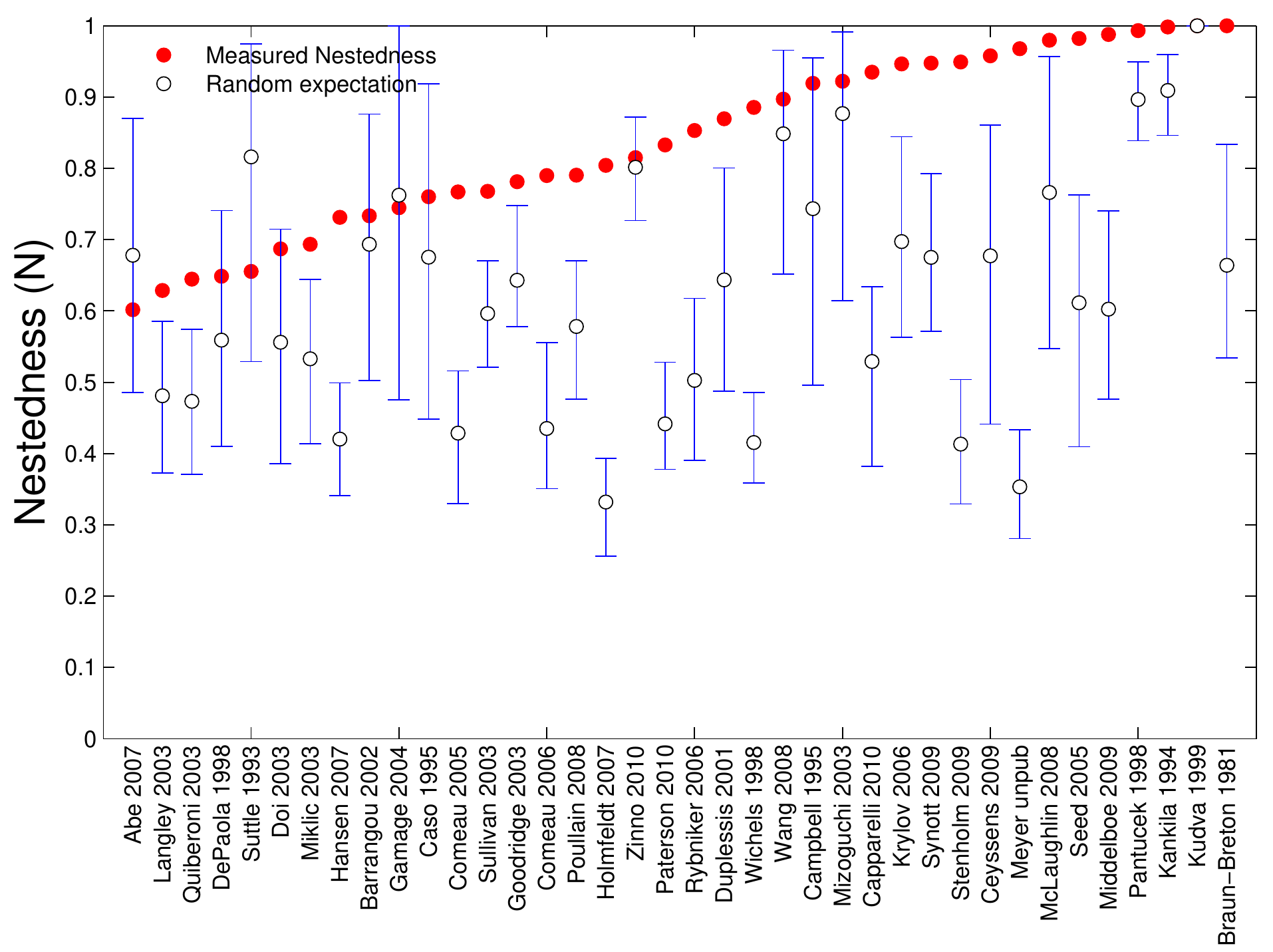}
 }
 \caption{Visual representation of the statistical tests in the set of matrices. Red circles
 represent the value of the analyzed networks. White circles represent the mean of the
 null model, while the error bars represent the networks that falls inside a two-tailed
 version of the random null model values. The margin of the error bars are (p-value,1-p-value),
 where p-value can is an optional argument of the plot functions.}
 \label{fig.random_exp}
\end{figure*}

First, the user should begin by creating an instance of
the \verb|MetaStatistics| class. This class takes, as input,
a cell array \verb|matrices| containing either a 
set of \matlab matrices or a set of
\verb|Bipartite| objects.  An automatic meta-analysis, using
default parameters, can be performed by the commands:
\begin{verbatim} 
mstat = MetaStatistics(matrices);
mstat.names = matrix_names %Labels for networks
%chosing the algorithms:
mstat.modularity_algorithm = @AdaptiveBrim
mstat.nestedness_algorithm = @NestednessNTC
mstat.DoMetaAnalyisis(); 
\end{verbatim} 
Results of the meta-analysis are stored in the object 
\verb|gstat|, for subsequent examination.  
The meta-analysis class (MetaStatistics.m) also has additional plot
functions. \textit{e.g.}, to compare network structures against
a null model values:
\begin{verbatim}
mstat.plotter.PlotModularValues(0.05);
mstat.plotter.PlotNestednessValues(0.05);
\end{verbatim}
where the argument represent the p-value threshold in determining
the variation about the network statistics generated from the
ensemble (lower values denote wider variation).
The output
for the modular and NTC values can be observed in Figure \ref{fig.random_exp}.
As is apparent, the majority of studies have modularity
\emph{below} that of the networks in the random ensemble.  In contrast,
the majority of studies have nestedness \emph{significantly above} that of
the networks in the random ensemble.

\begin{figure*}[t!]
	\centering
	\includegraphics[width=\textwidth]{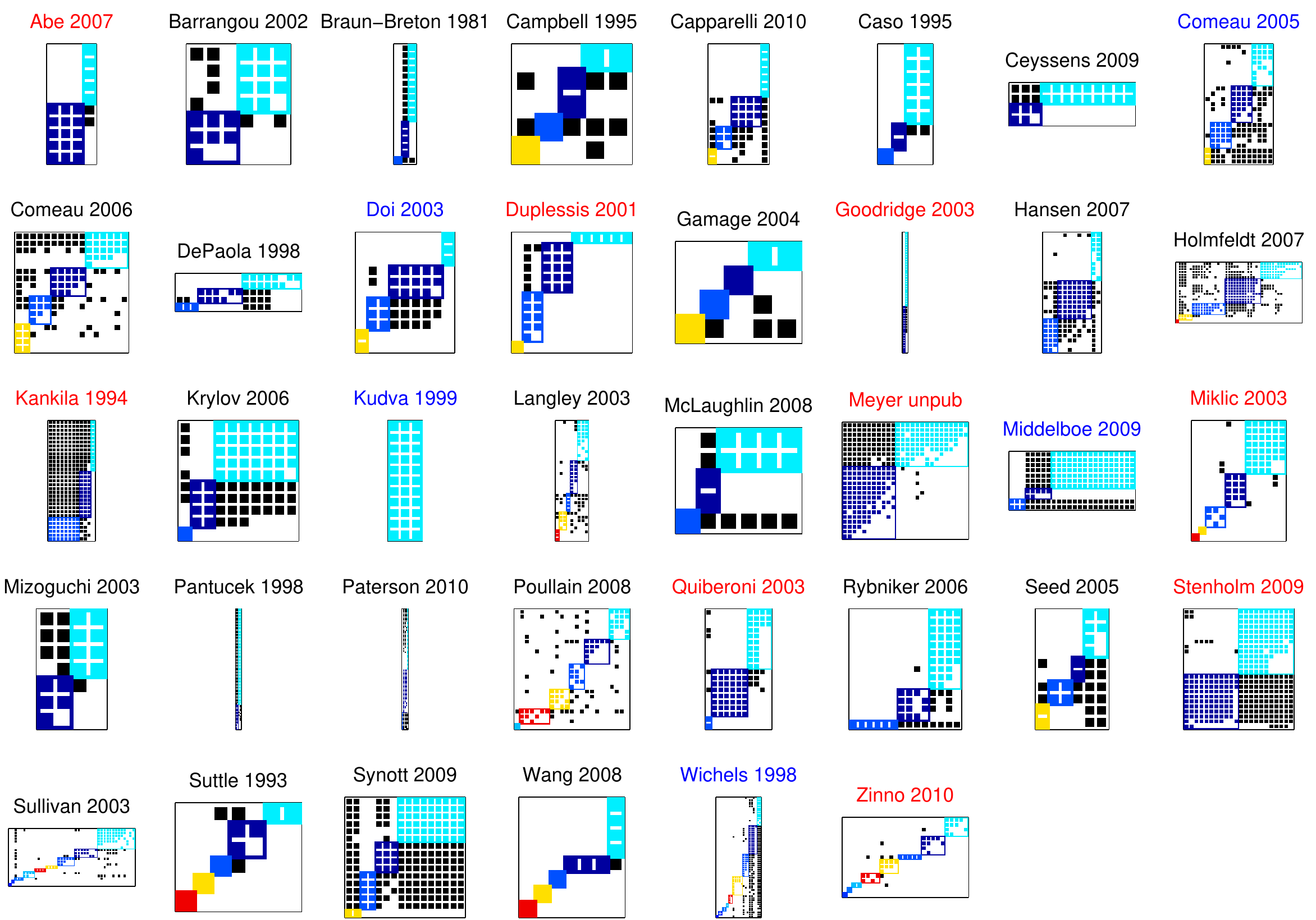}
	\caption{The meta-set collected on Flores et al \cite{flores2011statistical} plotted using the modularity
	algorithm of the \bimat library. Red and blue labels represent significant modularity ($p\ge 0.975$) and 
	anti-modularity ($p\le 0.275$), respectively. For bibliographic information about these matrices
	see \cite{flores2011statistical}.}
	\label{fig.modularity}
\end{figure*}

In addition, it is possible to plot all the matrices at once using any of
the next functions:
\begin{verbatim}
%Grid of 5 x 8
mstat.plotter.PlotMatrices(5,8);
mstat.plotter.PlotNestedMatrices(5,8,0.05);
mstat.plotter.PlotModularMatrices(5,8,0.05);
\end{verbatim}
where the first and second arguments are the number the matrices along
horizontal and vertical axis of the plot. If the statistical test have been
already performed, red and blue labels are used for indicate the statistical
significance of the corresponding structure (red for significance, and blue
for anti-significance), where the third argument (optional) is used to assess
a critical $p$-value
the significance. Figure \ref{fig.modularity} shows the plot for the case
of modularity.

\subsection{Example II: Multi-scale analysis}

\begin{figure*}[t!]
	\centering
	\includegraphics[width=\textwidth]{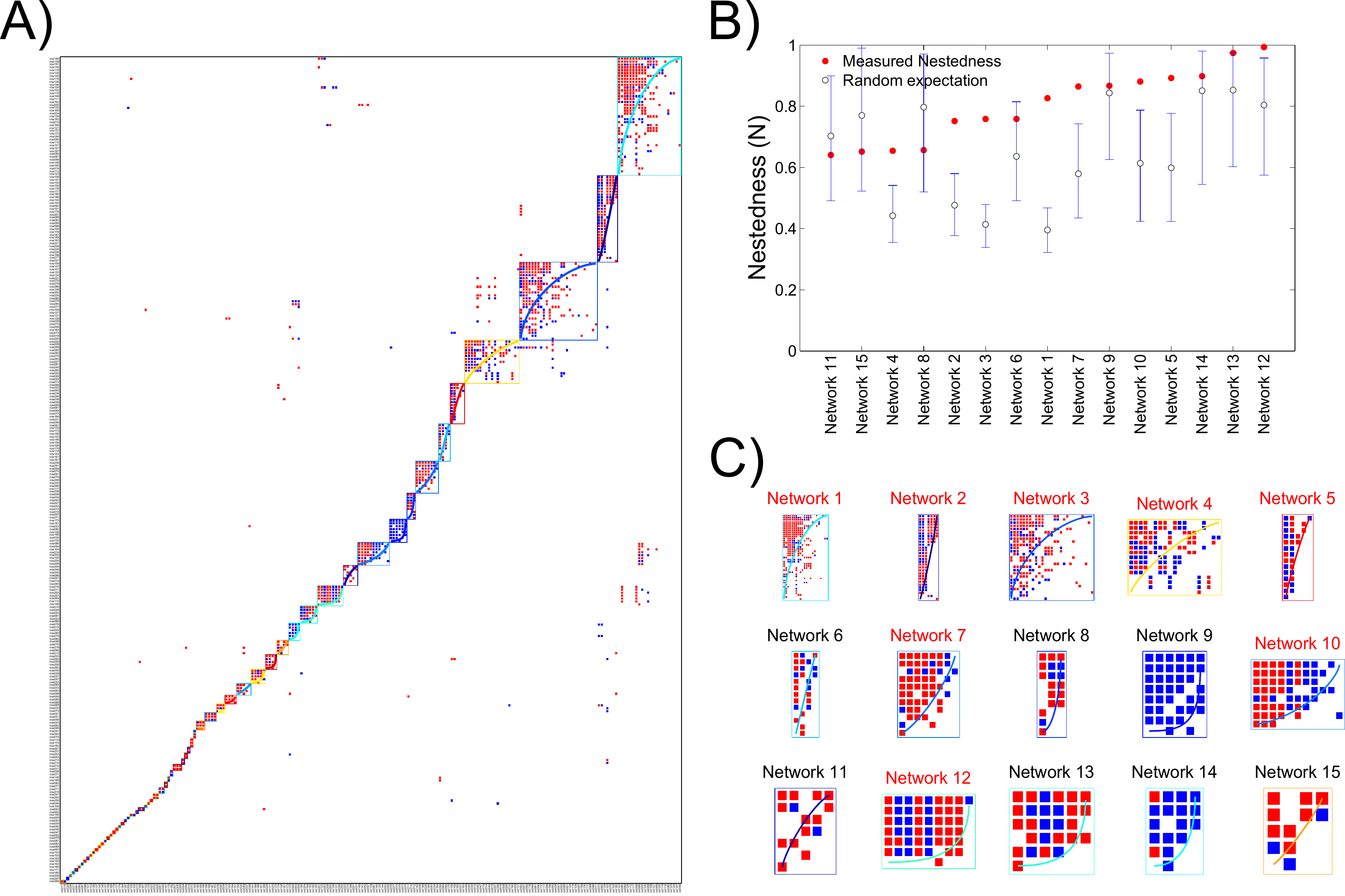}
	\caption{Standard plots that can be extracted using the multi-scale analysis capabilities
	of \bimat. Here, we focus in the internal nested structure using $N_{NTC}$ values, but we
	can also perform an internal study using $Q_b$ and $N_{NODF}$ values.
	\textbf{A)} The standard output using the modular matrix layout gives us a hint about
	the potential multi-scale structure. \textbf{B)} Here we focus on the study of $N_{NTC}$ values
	with respect to random expectation. Error bars cover 95 \% of the random replicate values.
	\textbf{C)} A more closer visual inspection on the analyzed matrices. Read labels indicate
	statistical significance of $N_{NTC}$ values.}
	\label{fig.moebus}
\end{figure*}

Moebus and Nattkemper \cite{moebus1981bacteriophage} published the largest
phage-bacteria infection network.
The individual phage and bacteria
were
extracted from different locations across the Atlantic Ocean. 
In a previous study we developed a multi-scale analysis
of network structure in this dataset\cite{flores2012multi}. 
Here, we demonstrate how such a multi-scale analysis can
be automated.
The first objective is to analyze the global-scale structure of a
bipartite network, i.e.
to quantify if the overall network has significantly elevated
or diminished modularity and/or nestedness.  Assuming that
our matrix is called \verb|moebus.weight_matrix| left panel of Figure~\ref{fig.moebus} shows
a visual representation of this data in matrix layout after typing:

\begin{verbatim}
bp = Bipartite(moebus.weight_matrix);
bp.community.Detect(); 
bp.plotter.font_size = 2.0;
figure(1);
bp.plotter.PlotModularMatrix();
\end{verbatim}

It becomes apparent that the network is modular. However, what is really
important to observe is that internal nodes seems to have nested structure 
(triangular pattern with most of the links above the
isocline). Hence, the Moebus network may have multi-scale structure properties.
We will confirm that this is the case for nestedness using the $N_{NTC}$ values. In order
to perform this test \bimat make use of the \verb|InternalStatistics| class in order to get the statistics
of those modules by isolating them and treating them as independent networks:

\begin{verbatim}
%We are interested in only the first 15 modules
%from the most righ-top one.
bp.internal_statistics.idx_to_focus_on = 1:15;
%Perform a default internal analysis
bp.internal_statistics.TestInternalModules();
figure(2);
bp.internal_statistics.meta_statistics...
             .plotter.PlotNestednessValues();
figure(3);
bp.internal_statistics.meta_statistics...
             .plotter.PlotNestedMatrices();
\end{verbatim}
where the last two plots are the ones on the right panels of Figure \ref{fig.moebus}.
The smart reader may already notice that \verb|meta_statistics| property is in fact an
instance of the class \verb|MetaStatistics|, which translates to be able to use any of the methods
inside \verb|MetaStatistics| (including its property \verb|plotter|) in the internal modules.

Finally, another multi-scale analysis that \bimat can perform is to quantify if a relation exist
between node classification and module distribution. If the extreme case, if this relation
exist nodes inside the same module will share the same classification. If the such
relationship does not exist, modules will have nodes with random classification. In order words,
the relationship depends in how random is the node classification inside the each module.
In order to perform this analysis \bimat make use of both Shannon's and Simpon's indexes.
And, for evaluating the significance we use a null model in which we randomly swap all node
classifications. We will give here a simple example about how to print the significance
of Simpson's index for the case of phage (column) nodes. In order to do so, we will use
geographical location extraction as classification identifier of each node:

\begin{verbatim}
% We want to use geographical location
% as classification
bp.col_class = moebus.phage_stations;
% Perform the analysis
bp.internal_statistics.TestDiversityColumns();
% Print results
bp.printer.PrintColumnModuleDiversity();
\end{verbatim}

The user must be able to visualize an output similar to:
\begin{verbatim}
Diversity index:    	  Diversity.SIMPSON_INDEX
Random permutations:	                      100
Module,index value,  zscore,percent
     1,    0.94805, -1.3848,      6
     2,    0.91738, -5.0054,      0
     3,    0.95238,-0.42625,     11
     4,    0.81667,-13.1025,      0
     5,          1, 0.36742,     12
     6,    0.85714, -2.5808,      0
     7,    0.66667, -2.4661,      0
     8,    0.33333,-13.5825,      0
     9,    0.90909, -2.0933,      3
    10,        0.9, -1.1203,      2
    11,        0.5, -6.6773,      0
    12,    0.88889, -2.6493,      1
    13,        0.6, -7.0097,      0
    14,        0.6, -8.0336,      0
    15,    0.83333, -1.3318,      3
\end{verbatim}

If we want to use the percentile as statistical test (using one-tail) and
$p$-value=0.5 we have that 12 modules are not as diverse as the random expectation.
Hence, these modules contain phages that come from similar geographical stations, which
translate to potentially have a relationship between the geographical location and module
formation for the phages case.

%\begin{figure}
%	\centering
%	\begin{subfigure}{0.58\textwidth}
%		\includegraphics[width=\textwidth]{matrix_modular}
%		\caption{Matrix modular layout}
%	\end{subfigure}
%	\begin{subfigure}{0.41\textwidth}
%		\includegraphics[width=\textwidth]{graph_modular}
%		\caption{Graph modular layout}
%	\end{subfigure}
%	\caption{Visual representation of the Moebus and Nattkemper \cite{moebus1981bacteriophage} bipartite
%	cross-infection network using two different layouts. \textbf{a)} Matrix layout: \bimat can color interaction cells if categorical data is used. Further, \bimat plot the modules in order to increase the visual aspect of their
%	nestedness. \textbf{b)} Graph layout: The same modular representation in a bipartite graph layout, where the left 
%	and right set represent bacteria and phage respectively.}
%\end{figure}

\clearpage

\section{Future Work}
We have developed \bimat -- an extensible \matlab library for the analysis of
bipartite networks.  \bimat implements standard algorithms
for the quantification of network structure,
including multiple tools to facilitate the analysis of the significance
of network structure at the whole network scale, across networks
and within networks.
The focus on two network features, modularity and nestedness,
reflects the importance both have in analyses of bipartite
network structure in ecological datasets.
However, these are not the only potential features of a bipartite network
nor are they necessarily independent.

Indeed, it has been suggested that modularity and nestedness
can be strongly correlated~\cite{fortuna2010nestedness}.
Such correlations may, on the one hand, lead to
spurious attempts at classifying a network as either
network or modular.  Poisot et al~\cite{poisot2012} have suggested
that bipartite networks may be classified based on the degree
to which a network is both
nestedness and modularity -- such classification may relate
to the presence of functional groups in the network.
Finally, both modularity and nestedness focus on structures
of the entire network.  However, non-random structures
may be present at alternative scales (\textit{e.g.}, see
the work on biological network motifs within
unipartite networks~\cite{alon1999broad}).
We have already made inroads in this direction
with a prior proposal~\cite{flores2012multi} and the current
automation of a multi-scale bipartite network analysis.
Future work is needed to evaluate the extent to which
the projection of bipartite networks into a lower dimensional
state space can help provide insights into distinct
types of networks and, eventually, on connections between network structure
and network function.

In moving forward, we hope that \bimat will become 
a dynamic, extensible tool of use to scientists
interested in bipartite networks.  
We are not the only group to propose such 
a comprehensive library.  For example, a team of UK scientists
recently proposed FALCON \cite{falconNest}, a library of tools
for the analysis of bipartite network structure in \matlab and R.  Similarly, we
are aware of unpublished efforts to develop a code-base
with similar toolsets in R\footnote{L.~Zaman, personal correspondence}.
The study of bipartite networks will necessarily involve
those with distinct scientific and computatoinal backgrounds.
Hence, so long as the code-bases are open-source, such efforts
are likely to reduce barriers in the analysis of bipartite
network structure, whether in the ecological, social 
or physical sciences.  

\section{Citation of methods implemented in \bimat}

The core algorithms implemented in \bimat are 
thoroughly described in their original publications
and discussed extensively by others.
In the case of nestedness, for the NTC
metric and implementation, see \cite{Atmar1993} and \cite{Rodriguez-Girones2006}
and for the NODF metric and implementation, see \cite{almeida2008consistent}. 
In the case of modularity, the standard BRIM
algorithms as well as its adaptive heuristic for module division are described by
\cite{Barber2007}. For a another heuristic using the standard BRIM algorithm, see
\cite{liuxin}. For the leading eigenvector algorithm, which is one of the most popular
algorithms in unipartite networks, see \cite{newman2006bmodularity}.

\section{Acknowledgments}
%We are indebted to people at the Sullivan's, Weitz's, and Acinas's groups, GEOMAR Helmholtz Centre for %Ocean Research, for the testing
%and feedback of this software.
COF acknowledges the support of
the CONACyT Foundation. 
TP thanks the FQRNT-MELS for funding through the PBEEE post-doctoral program.
JSW acknowledges NSF grant OCE-1233760 and support from
a Career Award at the Scientific Interface from the Burroughs Wellcome Fund.

\bibliographystyle{plain}
\bibliography{bimat}

\end{document}